 \documentclass{emulateapj}

\shorttitle{ICL at the Frontier: Abell 2744}
\shortauthors{M. Montes \& I. Trujillo } 

\def\gsim{ \lower .75ex \hbox{$\sim$} \llap{\raise .27ex \hbox{$>$}} }
\def\lsim{ \lower .75ex \hbox{$\sim$} \llap{\raise .27ex \hbox{$<$}} }

\begin{document}
\title{Intra-Cluster Light at the Frontier: Abell 2744}

\author{Mireia Montes$^{1,2}$ and Ignacio Trujillo$^{1,2}$} 
\affil{$^{1}$Instituto de Astrof\'{\i}sica de Canarias,c/ V\'{\i}a L\'actea s/n, E38205 - La Laguna, Tenerife, Spain}
\affil{$^{2}$Departamento de Astrof\'isica, Universidad de La
Laguna, E38205 La Laguna, Tenerife, Spain} 
\email{email:mireia.montes.quiles@gmail.com}

\begin{abstract}

The ultra-deep multiwavelength HST Frontier Fields coverage of the Abell Cluster 2744 is used to derive the stellar population properties of its intra-cluster light (ICL). The restframe colors of the ICL of this intermediate redshift ($z=0.3064$) massive cluster  are bluer ($g-r=0.68\pm0.04$; $i-J=0.56\pm0.01$) than those found in the stellar populations of its main galaxy members ($g-r=0.83\pm0.01$; $i-J=0.75\pm0.01$). Based on these colors, we derive the following mean metallicity $Z=0.018\pm0.007$ for the ICL. The ICL age is  $6\pm3$ Gyr younger than the average age of the most massive galaxies of the cluster. The fraction of stellar mass in the ICL component comprises at least $6\%$ of the total stellar mass of the galaxy cluster. Our data is consistent with a scenario where the bulk of the ICL of Abell 2744 has been formed relatively recently ($z<1$). The stellar population properties of the ICL suggest that this diffuse component is
mainly the result of the disruption of infalling galaxies with similar characteristics in mass
(M$_\star\sim 3\times10^{10}$ M$_{\odot}$) and metallicity than our own Milky Way. The amount of ICL mass
in the central  part of the cluster ($<400$ kpc) is equivalent to the disruption of $4-6$ Milky Way-type
galaxies.

\end{abstract}


\section{Introduction}

A substantial fraction of stars in clusters are not gravitationally bound to any particular galaxy. These
stars constitute the so-called intra-cluster light. The ICL is distributed around the central galaxy of
the cluster and extends to several hundred kpc away from the cluster center \citep[e.g.][]{Murante2004,
Zibetti2005}. This diffuse light is thought to form primarily by the tidal stripping of stars from
galaxies which interact and merge during the hierarchical accretion history of the cluster
\citep[e.g.][]{Gregg1998, Rudick2006, Conroy2007, Contini2014}. Therefore, the characterization of the ICL
provides a direct way of determining the assembly mechanisms occurring inside galaxy clusters.  In this
sense, the ICL is the signature of how violent the assembly of the cluster has been through its cosmic
history. For that reason, it is absolutely key determining how and when the ICL formed. However, the
identification of this light observationally remains difficult and uncertain. Indeed, the typical surface
brightness of the ICL is $\mu_{V} \gtrsim 26.5$ mag/arcsec$^2$ \citep[e.g.][]{Mihos2005,
Zibetti2005,Rudick2006} and it is contaminated by foreground and background galaxies. Moreover, it is
difficult to dissociate between the ICL and the brightest central galaxy surface brightness profile
\citep[e.g.][]{Gonzalez2005, Krick2007}.

Although the properties of the stellar populations of the ICL provide a vital tool to understand its
formation, little is known about their characteristics. The ICL is reported to be metal-poor in some
studies \citep[e.g.][]{Durrell2002, Williams2007} while other studies show super-solar metallicities
\citep[e.g.][]{Krick2006}. Simulations predict that the ICL forms very late \citep[$z<1$;
e.g.][]{Murante2007, Contini2014} and that the bulk of the ICL is produced by the most massive
(M$_\star\sim10^{10-11}$ M$_{\odot}$) satellites as they fall into the cluster core
\citep[e.g.][]{Murante2007, Purcell2007, Martel2012, Contini2014}. If this scenario is correct, the ages
and metallicities of the intracluster population will be highly dependent on the progenitor galaxies from
which they were stripped. Therefore, the ICL is expected to have a metallicity that is similar to that of
these massive satellites. The goal of this paper is to explore this question in detail and characterize, for
the first time, the age and metallicity of the ICL of a massive cluster at radial distances R$>50$ kpc
with unprecedented accuracy.

We achieve our goal taking advantage of the deepest optical and near-infrared images ever taken by
the Hubble Space Telescope (\emph{HST}) of the  Abell Cluster 2744. The Hubble Frontier
Fields\footnote{http://www.stsci.edu/hst/campaigns/frontier-fields} (HFF) project represents the largest
investment of HST time for deep observations of galaxy clusters. Abell 2744 is the first cluster observed
as part of the HFF program. This structure, at $z=0.3064$ \citep{Owers2011}, is a rich cluster (virial
mass of $\sim7\times10^{15}$ M$_{\odot}$ within $R<3.7$ Mpc) undergoing a major merger as evidenced by its
complex internal structure \citep{Boschin2006}. Therefore, this cluster represents an excellent target for
studying the formation of the ICL. The ages and metallicities of the ICL are studied in detail taking
 advantage of the inclusion of very deep near-infrared (NIR) data to break the age-metallicity
degeneracy \citep[e.g.][]{Anders2004}.

Throughout this work, we adopt a standard cosmological model with the following parameters: $H_0$=70 km s$^{-1}$
Mpc$^{-1}$, $\Omega_m$=0.3 and $\Omega_\Lambda$=0.7. At $z=0.3064$, this corresponds to a spatial scale of $4.52$ kpc/arcsec. 

\section{Data}\label{data}

The data used in this work are based on the first release of the HFF program and include all the NIR WFC3
data (ID13495, PI: J. Lotz and ID13386, PI: S. Rodney) of  Abell 2744. The ACS images were taken from the
HST archive as part of the program: ID11689 (PI: R. Dupke) and consist of six orbits in F435W and five
orbits each in F606W and F814W. NIR observations include imaging in four filters F105W, F125W, F140W,
F160W based on 24, 12, 10, and 24 orbits, respectively. The data were directly retrieved from the
archive\footnote{http://www.stsci.edu/hst/campaigns/frontier-fields/FF-Data}. The images were reduced by
the HFF team following standard HST procedures both for the ACS and WFC3 data\footnote{More details on the
FF images reduction in:\\
http://archive.stsci.edu/pub/hlsp/frontier/abell2744/images/
hst/v1.0/hlsp\_frontier\_hst
\_wfc3-acs\_abell2744\_v1.0\_readme.pdf}.
For both cameras, flat fields are claimed to be accurate to better than $1\%$ across the detector. A few of the HFF observations in the IR exhibit a time-variable sky background signal due to time variable atmospheric emission. This exposures were corrected from this variable emission and included into the final mosaics (Koekemoer 2014, priv. comm.). 

The mosaics we used consist on drizzled science images with pixel size $0\farcs06$. These mosaics were
build using the package Astrodrizzle\footnote{http://drizzlepac.stsci.edu/}. In the case of the WFC3, 
this pixel size is closer to one half of an original pixel. While working with subpixel dithered data has
several advantages as helping to the comic ray rejection step, improving the resolution and lowering the
correlated noise, it also reduces the sensitivity to low surface brightness features. This last problem,
however, can be avoided by combining several pixels later on as we have done in our work (see next
section).

The $0.4$\arcsec diameter aperture depth at $5\sigma$ of each image is: $27.4$ (F435W), $28.0$ (F606W), $27.1$ (F814W), $28.6$ (F105W), $28.5$ (F125W), $28.7$ (F140W) and $28.2$ (F160W) mag \citep{Laporte2014}.
As we will explain later, the age and metallicity of the ICL is determined  using the color information provided by both cameras: ACS and WFC3. Consequently, we will limit our study of the ICL to their common field of view ($\sim2.0\times2.2$ arcmin$^2$) of the galaxy cluster.

\subsection{Surface brightness limits}

Our goal is to study the properties of the stellar populations of Abell 2744 down to the faintest surface
brightness possible. For this reason, it is necessary to measure accurately the colors of the stellar
populations across the cluster. Given that the ACS images are shallower than the WFC3 data, we have
spatially re-binned the images of the HFF to improve the S/N and to maximize the detection of the diffuse
light of the cluster. Previous to the rebinning, we have smoothed the images with a $5$ pixel median box
filter to reduce noise, specially in the outer regions of the cluster. The new rebinned pixel value is the
sum of $25$ input pixels in the original image and has a size of $0.3$\arcsec ($\sim1.4$ kpc at z$=0.3064$). The process of smoothing and rebinning the images is flux conserving.

We have estimated the surface brightness limits of our images as follows. We have identified $20$ background
regions spread over the outer parts of the images. The background regions are chosen to avoid the contamination of the light of the cluster. With this aim we have selected areas with $\mu_J>28$ mag/arcsec$^2$, $\rho<10^{0.7}$ M$_\odot$/pc$^2$ and $R>230$ kpc (see Section \ref{parameters}). We have explored whether the light distribution in the background area follows a normal distribution. A Kolmogorov-Smirnov test indicates that a gaussian distribution is compatible with the data (the null hypothesis is not rejected with any statistical significance). We do not find any evidence supporting that our background distributions present gradients.

To estimate the surface brightness limits, we calculated the r.m.s of the images on boxes of $3\times3$ arcsec$^2$ (equivalent to $10\times10$ kpc$^2$ at the cluster redshift). The surface brightness limits we provide correspond to $3\sigma$ detections. These limits are:  $29.31\pm 0.02$ (F435W), $29.60\pm 0.10$ (F606W), $29.13\pm0.07$ (F814W), $30.32\pm0.08$ (F105W), $29.97\pm0.09$ (F125W), $30.01\pm0.05$ (F140W) and $30.05\pm 0.03$ (F160W) mag/arcsec$^2$. 
Although the HFF data is sky substracted, in order to have a better sky determination, we have used the above regions to reestimate the background of each of the images and subtracted (added) it to the whole image.

Finally, for the analysis of the ICL, it is crucial to identify the galaxies that are members of the cluster and
get rid of background and foreground sources. For this purpose, we build a redshift mask of the cluster
image. The mask was constructed using the spectroscopic redshift catalog of \citet{Owers2011} and adding
all the photometric redshifts available in NED\footnote{http://ned.ipac.caltech.edu/}. To assign a
redshift value to the pixels of the images, first we run \textsc{sextractor} in the F160W image to
identify the area that correspond to each object, using a detection threshold at a 5$\sigma$ significance level and a gaussian convolution kernel with size $0.75$\arcsec. Then, we masked the area subtended by those objects whose redshifts do not correspond to the cluster ($z>0.33$ and $z<0.27$) and also those without measured redshifts. 

\subsubsection{Background level uncertainties}\label{blu}

The uncertainties at determining the surface brightness limits of the images reflect how accurate we can estimate the background level. We have quantified how these uncertainties affect the determination of the fluxes in the ICL region of our images. As we will see later, we define the ICL as the region of the cluster with restframe surface brightness 24$<\mu_J<$25 mag/arcsec$^2$. Placing in this region a $3\times3$ arcsec$^2$ box, we measured the flux in it. This ICL flux was then compared with the uncertainties at the determination of the background. We find that the previous background uncertainties translate into the following errors on the different bands fluxes in our ICL bin: $24$\% (F435W), $5.3$\% (F606W), $4.1$\%(F814W), $0.4$\% (F105W), $0.3$\% (F125W), $0.4$\% (F140W) and $0.1$\% (F160W). As expected, the errors are larger for our ACS filters, which have shallower exposures, than for our WFC3 filters. However, it is worth noting, that our work only uses data based on bands redder than F606W. Consequently, the uncertainties at measuring the background level only affect the fluxes in the ICL region below 5\% in all cases. This little effect, particularly in the NIR bands, is expected because our surface brightness limit to study the ICL is well above ($\sim4$ mag) the limiting surface brightness of our imaging. This guarantee that we are working in a surface brightness regime less affected by problems of flat fielding, incorrect sky subtraction, etc. In Table \ref{table2}, we summarize the sources of noise that can affect our measurements.

 \begin{table*}
  \centering
  \tabcolsep 2.5pt
   \begin{tabular}{@{}cccc@{}}
    \hline
    Sources of noise             &     Affecting            &   Relative impact             & Addressed in Section \\ \hline
    Time variable sky background&  Few exposures in the IR & Modeled during the reduction process  & \ref{data} \\
    Flat-field	accuracy	&  All images 		   & $<1\%$ across the detector   & \ref{data} \& \ref{results}\\
    PSF uncertainties		&  All images		   & $g-r<0.03$ \& $i-J<0.02$      & Appendix  \\
    Background estimation 	&  All images		   & No gradients. Errors of $<5\%$&  \ref{blu}\\ \hline
   \end{tabular} 
  
  \caption{Summary of the sources of noise/uncertainty present in our analysis.}\label{table2}

 \end{table*}

\subsection{Optical and NIR colors}\label{parameters}

In order to study the stellar populations of the cluster, including its ICL, we constructed two restframe colors
at the redshift of the cluster: $g-r$ and $i-J$. The choice of these colors was made for two reasons: first, to
diminish the effects of the different PSFs combining images from the same camera (ACS data for $g-r$ and WFC3
data for $i-J$) and second, to include a NIR filter to constrain the metallicity  with better accuracy
\citep[e.g.][]{Anders2004}. The colors  were built interpolating among the observed filters to obtain the flux
in the equivalent restframe bands: $g$, $r$, $i$ and $J$. We also derived the $z$ band, using the same methodology, to construct the $r-z$ color. This color was used to create a measurement of the stellar mass density of the cluster (see below). We corrected our data of Galactic extinction (E(B-V)=$0.012$, \citealt{Schlafly2011}) using a \citet{Cardelli1989} extinction law.

To explore the stellar populations of Abell 2744 and their variation across the cluster structure, we make
use of three independent analysis based on different parameters: the restframe (corrected of cosmological
dimming) surface brightness in $J$-band, $\mu_J$, the logarithm of the stellar mass density, $log(\rho)$,
and the radial distance, $R$, to the most massive galaxies of the cluster. These three parameters will allow
us to compare our results with theoretical expectations.

The $\mu_J$ map of the cluster was divided in eight surface brightness bins, from $16$ to $25$ mag/arcsec$^2$. To
estimate $log(\rho)$, we follow the procedure described in \citet{Bakos2008}. First, we link the stellar mass density  profile $\rho$ with the surface brightness profile at a given wavelength $\mu_{\lambda}$ using the mass-to-light (M/L) ratio. This is done with the expression: 

\begin{equation}
 \log(\rho)=\log(M/L)_{\lambda}-0.4(\mu_{\lambda}-m_{abs\, \odot \lambda})+8.629
\end{equation}

where $m_{abs\, \odot \lambda}$ is the absolute magnitude of the Sun at wavelength $\lambda$ and $\rho$ is measured in M$_{\odot}$ pc$^{-2}$. Second, to evaluate the above expression, we need to obtain the M/L ratio at each radius. Following the prescription of \citet{Bell2003}, we have calculated the M/L ratio as a function of color. In this work
we have used the restframe color $r-z$ and assumed a Salpeter IMF \citep{Salpeter1955}. The expression we
have used to estimate the M/L is:

\begin{equation}
\log(M/L)_{\lambda}=(a_{\lambda }+b_{\lambda }\times color)
\end{equation}

where $color=r-z$,  $a_{\lambda}$=-0.041 and $b_{\lambda}$=0.463 are applied to determine the M/L in the
z-band. The $log(\rho)$ map of the cluster was divided also in eight mass density bins, from $10^{4.8}$ to $10^{0.8}$ M$_{\odot}$/pc$^2$. 

\begin{figure*}
 \includegraphics[scale=0.8]{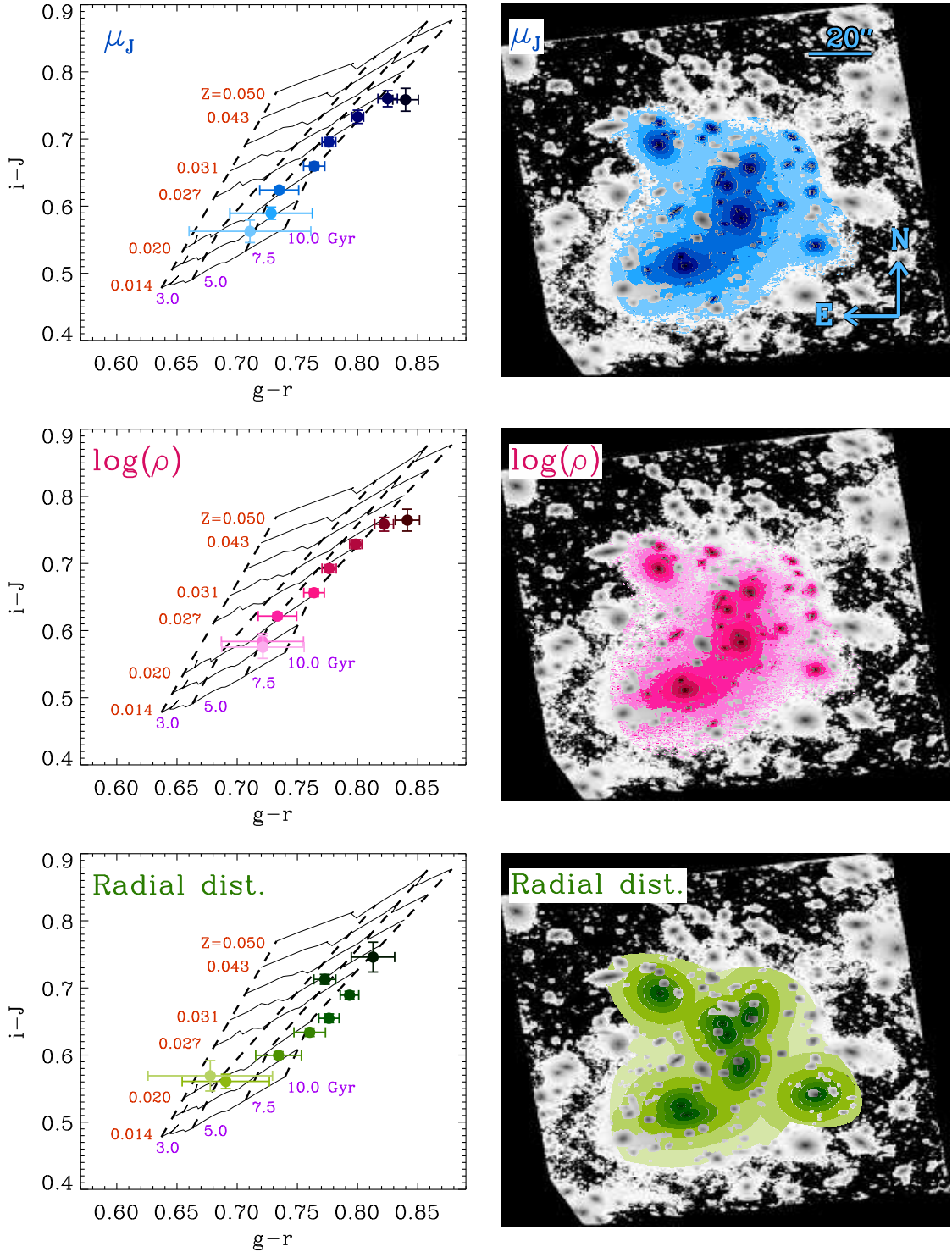}
 
 \caption{Left panels show the $i-J$ vs. $g-r$ diagrams for  three different parameters: $\mu_J$ (blue),
$\log(\rho)$ (pink) and $R$ (green) used  to characterize the different spatial regions where the light of
the cluster is distributed. Overplotted is a color grid prediction based on \citet{Bruzual2003} models
using a Salpeter IMF for single stellar populations. Right panels show the image of the cluster in the
F160W filter and over-plotted are the different spatial regions in which the colors are measured. The
spatial regions are coded from the inner parts of the cluster (darker color) to the outer parts (lighter
color).}

 \label{fig1}
\end{figure*}

Finally, we constructed a \emph{radial distance} indicator using the centers of the most massive galaxies
in the cluster as the starting points. These massive galaxies were identified as those galaxies with
central surface brightness $\mu_J<17$ mag/arcsec$^2$. This surface brightness translates to a density of $10^4$ M$_\odot$/pc$^2$ equivalent to the stellar mass densities in the centers of massive ellipticals \citep[e.g.][]{Hopkins2009}. Then, the distance to their centers was calculated as the elliptical distance defined by the morphological parameters of these galaxies, derived by \textsc{sextractor}. The less massive galaxies of the cluster were masked to reduce the contamination of their different stellar populations at a given radius. This radial distance was logarithmic spaced in eight bins, from $0$ to $120$ kpc. We illustrate the different bins in Fig. \ref{fig1}.

The reason to choose the above parameters, and their binning, is to characterize the stellar populations of the
cluster properly, i.e. averaging zones with similar properties. This allows us to describe the cluster from its
inner parts (brighter magnitudes, higher densities) to its outer parts (fainter magnitudes, lower densities) in
a consistent way. On doing this, the comparison of the ICL properties with those from the cluster galaxies is
direct and homogeneous. Consequently, we will infer the properties of the ICL and cluster galaxies based on
their relative differences. 

In Fig. \ref{fig1}, the three color-color diagrams based on the $\mu_J$, $log(\rho)$ and $R$ maps of the
cluster are shown (left panels). A grid of \citet{Bruzual2003} models based on a Salpeter IMF for single
stellar populations is also plotted. Each of the points of the grid is flagged with its corresponding age
and metallicity. For illustration, in the right panels, the color-coded bins of the three parameters are
drawn over an image of the cluster in the F160W band.

\section{Results}\label{results}

%%%%%%%%%%%%%%%%%%%%%%%%%%%%
%%%%%%%%%%%%%%%%%%%%%%%%%%

\subsection{Color-color diagrams}

The left panels of Fig. \ref{fig1} show the color-color diagrams for each parameter used to characterize
the cluster light, i.e. $\mu_J$, $\log(\rho)$ and $R$. Each data point on the color-color diagram
corresponds to the average restframe $g-r$ and $i-J$ color values of the spatial regions drawn in the cluster maps
plotted in the right panels. The errors in the $g-r$ and $i-J$ colors are derived from bootstrapping
simulations. These simulations consist of $3000$ realizations randomly choosing half of the pixels in each
bin and calculating their mean colors. The errors represent the scatter of the distributions of the means.
These errors account for both the intrinsic scatter of the cluster's stellar populations and the
photometric errors. In addition, this bootstrapping technique could also account for local variations in the flat-field of the images.

For all the color-color diagrams, a continuous bluing of both $g-r$ and $i-J$ colors is clearly seen.
Therefore, the stellar populations of the cluster (galaxies and ICL) become gradually bluer when moving
away from the centers of the galaxies (i.e., towards fainter luminosities and lower densities). According
to the values of age and metallicity indicated by the grid of models, this gradient in colors is
compatible with a gradient in metallicity, from a supersolar metallicity to a little subsolar, and also a
slight gradient in age (see next subsection). It is worth noting a slightly discrepant point ($\lesssim3 \sigma$) at $\sim5$ kpc in the radial distance plot. We think this discrepant point is due to a combination of low statistics and mixing of different stellar populations at that radius.

A potential source of concern in our color determination is the effect of the PSF on the color
distribution. Consequently, it is important to quantify a possible contamination of the colors in the
outer regions of the cluster due to the scattered light produced by the different band PSFs. We have
quantified this effect by modeling the effect of the different band PSFs in a model galaxy without and with a modeled ICL component. This is described in the Appendix. We find that the effect of the PSF both for the $g-r$ and $i-J$ colors is small
(less than $0.03$ and $0.02$ mag, respectively, at the region of the ICL, i.e. $R>50$ kpc) and less than $0.01$ mag when adding an ICL component. Moreover, if any, the PSF effect on the color profile tends to redden the color. As we are observing a continuous bluing of the color radial profile, we conclude that the effect of the PSF plays a minor role in our analysis.

\subsection{Age and metallicity gradients}

\begin{figure*}[!h]
 \begin{center}
  \includegraphics[scale=0.5]{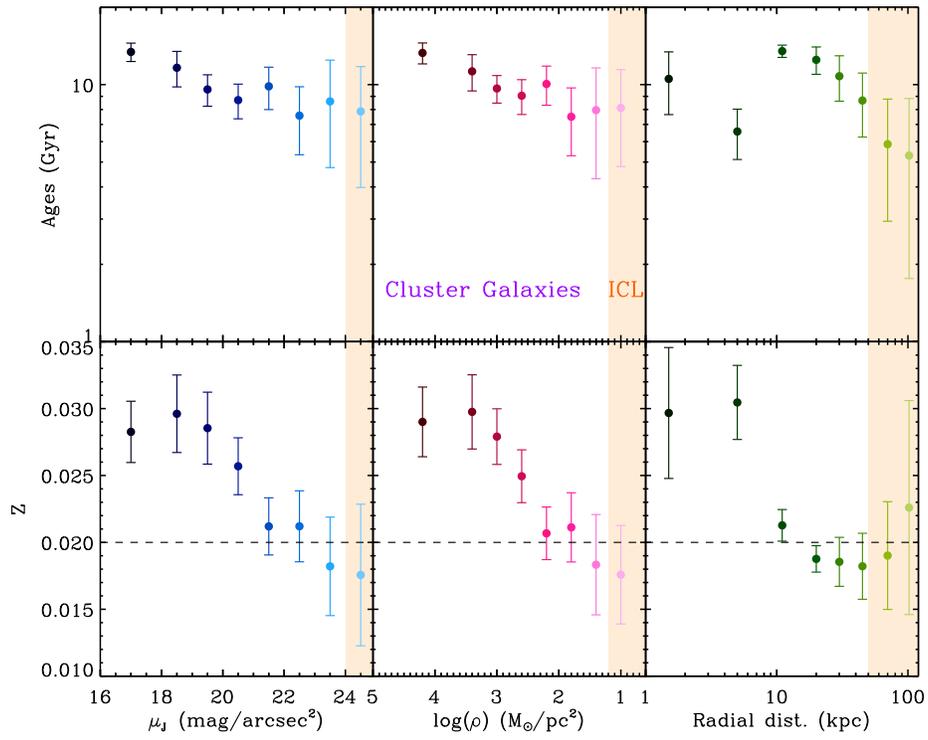}
  
  \caption{Gradients of age and metallicity as a function of $\mu_J$, $\log(\rho)$ and $R$. The regions
corresponding to the cluster galaxies (purple) and the ICL (orange) are labeled. The dashed line indicates
the solar metallicity.} 

 \label{fig2}
 \end{center}
\end{figure*}

To quantify the changes of the stellar populations across the galaxy cluster, we have computed the age and
metallicity of the stellar populations based on their restframe $g-r$ and $i-J$ colors. Within each spatial bin, we
measure the age and the metallicity of each pixel corresponding to their $g-r$ and $i-J$ colors. Then, we
estimate the corresponding mean age and metallicity. The errors in age and metallicity were also drawn from the bootstrapping simulations used to derive the color errors. The gradients of age (upper panels) and metallicity (bottom panels) for each of the parameters are shown in Fig. \ref{fig2}. The colors, ages and metallicities for the three parameters are listed in Table
\ref{table1}.

%%%%%%%%%%%%%%%%
%%%%%%%%%%%%%%%%

In Fig. \ref{fig2}, tentatively, we marked in orange the region corresponding to the ICL component.
Observationally, the exact definition of which part of the cluster light describes the ICL is ill-defined.
There have been many different attempts to label the ICL region using criteria as the surface brightness
or the radial distance. For instance, \citet{Zibetti2005} defined the ICL as the light with $\mu_r>25.0$
mag/arcsec$^2$ whereas \citet{Toledo2011} used $R>50$ kpc \citep[see also][]{Gonzalez2005}. For each of the parameters used, we have defined the ICL region as the region with either restframe $\mu_J>24$ mag/arcsec$^2$, $log(\rho)<1.2$ or $R>50$ kpc. The $\mu_J$ limit defining the ICL region is based on the Zibetti et al. definition and was calculated using $i-J=0.6$ (see Fig. \ref{fig1}) and $r-i=0.4$ \citep[][for solar metallicity and $\sim7$ Gyr]{Vazdekis2012} while the $log(\rho)$ limit is calculated using the above $\mu_J$ limit and the color $i-J=0.6$. These values are inserted into equation 1 in \citet{Bakos2008} with the mass to light ratio in the $J$-band given by \citet{Vazdekis2012} models for the corresponding age ($\sim7$ Gyr) and metallicity ($Z\sim0.018$) of the ICL region.

According to Fig. \ref{fig2}, the metallicity within the cluster is continuously decreasing from
supersolar ($Z=0.028\pm0.003$) in the central part of the main galaxies to solar ($Z=0.018\pm0.007$) in the ICL region. The ICL metallicity is similar to that found in the stellar halo of nearby massive galaxies \citep{Coccato2010, Roediger2011, Greene2012, Montes2014}. The stellar population ages show a slight negative gradient towards the outskirts. This gradient is compatible with what is found for the giant elliptical galaxy M87 \citep{Liu2005, Montes2014}, the elliptical galaxies in \citet{Roediger2011} and for some galaxies in \citet{Greene2012}.

\section{Discussion}

State-of-the-art semianalytical models of galaxy formation \citep[i.e.][]{Contini2014} suggest that the origin of the ICL is the result of the disruption and tidal stripping of massive ($10^{10-11}$ M$_{\odot}$) satellite galaxies infalling in the cluster potential \citep[see also][]{Purcell2007, Murante2007, Martel2012}. The largest contribution of these massive satellites to the ICL is understood due to the stronger effect of dynamical friction on these galaxies orbiting the cluster compared to less massive satellites. If this theoretical scenario is correct, we would expect that the mean metallicity of
the ICL would be slightly subsolar Z$\sim0.009-0.014$ (\citealt[]{Contini2014}, corresponding to the typical metallicities of the galaxies described above). Here we find that the mean metallicity of the ICL in our cluster is solar ($Z=0.018\pm0.007$), but still in agreement with the theoretical predictions. Using the mass-metallicity relation \citep{Gallazzi2005}, the derived ICL metallicity corresponds to the metallicities of galaxies with M$_\star\sim$ 3$\times$10$^{10}$ M$_{\odot}$. This mass is only a factor of two different to that of the Milky Way and the metallicity is similar ($6.43 \pm0.63 \times10^{10}$ M$_{\odot}$, \citealt{Mcmillan2014}; $Z\approx0.02=Z_{\odot}$, \citealt{Rix2013}). Therefore, the ICL of Abell 2744 can be understood as  mainly produced by the disruption of galaxies with similar stellar properties as the Milky Way.

Many theoretical works \citep[e.g.][]{Willman2004, Monaco2006, Murante2007} find that the majority of the ICL formed at $z<1$. According to \citet{Contini2014}, it is only since $z=0.4-0.5$ that $50\%$ of the ICL of present-day clusters is in place. We find that the ICL age of Abell 2744 is younger ($6\pm3$ Gyr) than the age of the most massive (M$_{\star}\gtrsim10^{11}$ M$_{\odot}$) galaxies of the cluster. This is consistent with the idea that the galaxies that mostly contributed to the ICL were producing stars during a larger period of time than the most massive galaxies of the cluster. This could happen if the satellite galaxies orbiting the cluster were forming stars until their star formation ceased due to ram pressure stripping of their gas content \citep[e.g.][]{Boselli2009, Chung2009}. Later on, with their star formation stopped, the stellar populations of these satellite galaxies could be stripped from their progenitors and started to contribute to the ICL. However, it is not straightforward to say exactly when the ICL in Abell 2744 was produced. Nonetheless, we can speculate on the basis of the difference between the age of the most massive galaxies and the one derived for the ICL\footnote{We do not use the absolute value of the ages of the stellar populations as it is well known that these values are more prone to errors than their relative differences \citep[e.g.][]{Vazdekis2001}.}. If we assume that the stellar populations of the most
massive galaxies were formed at z$\sim2-3$ (i.e. $\sim10-11$ Gyr ago), then the ICL of Abell 2744 is compatible with being assembled at $z<1$. 

The restframe optical color of the ICL of Abell 2744 ($g-r\sim0.68\pm0.04$) is consistent with the results of
\citet{Krick2007} for the ICL of the same cluster ($V-r=1.0\pm0.8$). We also compare the color of the ICL
of Abell 2744 with the ICL colors available for other clusters. As mentioned before, there are few studies
regarding the properties of the ICL, even colors. We warn the reader that the clusters studied in the
literature have different masses and redshifts making the comparison of such properties not
straightforward. The derived color for the ICL of Abell 2744 is in agreement with the colors derived for
various tidal features in the Virgo Cluster ($B-V\sim 0.75$, \citealt{Rudick2010}) and the outskirts of
M87 ($g-r\approx0.72$, \citealt{Chen2010}, priv. comm.). \citet{Covone2006} also found blue colors
($B-V\sim1.0\pm0.5$) for two diffuse features in the cluster Abell 2667 at $z=0.233$ compared to the colors of
the most massive galaxies ($B-V\sim1.5$), although there is a hint of a redder $V-I$ color for the ICL ($V-I\sim1.0\pm0.3$ for the ICL, $V-I\sim0.7$ for the galaxies). 

Finally, another aspect of the ICL that can be directly compared with the theoretical predictions is the
fraction of stellar mass of the galaxy cluster that is contained within this component. As we mentioned
earlier, the precise definition of what is exactly ICL is observationally ill-defined. Here we have used
up to three independent parameters to quantify the spatial region corresponding to the ICL:  $24<\mu_J<25$
mag/arcsec$^2$, $0.8<log(\rho)<1.2$ and $50<R<120$ kpc. The amount of stellar mass in the spatial regions defined
by those constraints were derived as follows. For $log(\rho)$ the derivation of the mass was
straightforward, while for $\mu_J$ and the radial distance we used Equation 1 and the corresponding mass to light ratio in the $J$-band for the age and metallicity of the ICL region (see Fig. \ref{fig2}). The total mass of the cluster, including the ICL region, was obtained similarly but using all the spatial region covered by the light of the cluster explored in this work (i.e. $\sim400\times400$ kpc around its core). The amount of total stellar mass in this region of the cluster is: $7.5\times10^{12}$ M$_\sun$ ($\mu_J<25$ mag/arcsec$^2$), $7.9\times$10$^{12}$ M$_\sun$ ($\log\rho>0.8$)
and $4.2\times10^{12}$ M$_\sun$ ($R<120$ kpc, using our definition of radial distance to the centers of the most massive galaxies of the cluster). The ICL mass fractions we got are: $5.1\%$ for $\mu_J$, $4.0\%$ for $log(\rho)$ and $10.4\%$ for the radial distance. These ICL mass fractions correspond to the mass contained in $4-6$ Milky Way-type galaxies. Note, however, that these fractions are lower limits to the global contribution of the ICL to the total stellar mass of the cluster. This is due to the fact that we are not considering the whole cluster (but only $\sim400\times400$ kpc around its core) and we are also limited by the depth of the optical bands (particularly for estimating $log(\rho)$ based on the $r-z$
color). How our numbers compare with the theoretical expectations? Simulations predict that the fraction
of mass in the ICL of present-day galaxy clusters should be around $10\%$-$40\%$ \citep{Contini2014}. At
the redshift of our cluster, $z=0.3$, the fraction of stellar mass in the ICL is expected to be around
$60\%$ of today's value. Consequently, we assume that the fraction of mass in a cluster as Abell 2744
should be between $6\%$-$24\%$. Taking into account that our stellar mass estimates are lower limits, our
numbers are in good agreement with theoretical expectations. The simulations are also in nice agreement
with other stellar mass fractions measured in other clusters ($10\%$-$20\%$; \citealt{Feldmeier2004,
Covone2006, Krick2006, Krick2007}). Over and over, it seems that the theoretical predictions are
reproducing well the observations presented in this work and in other previous papers. We think this is an
indication that we are starting to understand the general picture of how the ICL of the galaxy clusters
form.

The results presented in this work show the extraordinary power of the Frontier Fields survey to address
the origin and evolution of the ICL. Once this survey will be completed, it will be possible to explore the
properties of the ICL in another $5$ clusters in the redshift range $0.3-0.6$; a period of time crucial to
understand the formation of this elusive component of the galaxy clusters.

\acknowledgments  

We thank the referee for constructive comments that helped to improve the original manuscript and the HST
director and FF teams for their work to make these extraordinary data available. We also thank A. M. Koekemoer for his useful comments in the HFF image reduction process. This research has been supported by the Spanish Ministerio de Econom\'ia y Competitividad (MINECO; grant AYA2010-21322-C03-02). 
\\

\begin{table*}
 \centering
  \tiny
  \tabcolsep 2.5pt
   \begin{tabular}{@{}ccccc@{}}
    \hline
    \multicolumn{5}{c}{$\mu_J$ (mag/arcsec$^2$)}  \\ \hline
 Bin   &$g-r$ & $i-J$ & Age (Gyr) & Z \\
16-18 &$ 0.840\pm 0.011$&$ 0.759\pm 0.017$&$13.401\pm 1.102$&$0.028\pm0.002 $\\
18-19 &$ 0.825\pm 0.008$&$ 0.760\pm 0.012$&$11.631\pm 1.835$&$0.030\pm0.003 $\\
19-20 &$ 0.800\pm 0.005$&$ 0.733\pm 0.010$&$ 9.579\pm 1.339$&$0.029\pm0.003 $\\
20-21 &$ 0.776\pm 0.006$&$ 0.695\pm 0.007$&$ 8.703\pm 1.341$&$0.026\pm0.002 $\\
21-22 &$ 0.764\pm 0.009$&$ 0.660\pm 0.006$&$ 9.846\pm 1.847$&$0.021\pm0.002 $\\
22-23 &$ 0.735\pm 0.016$&$ 0.624\pm 0.004$&$ 7.574\pm 2.238$&$0.021\pm0.003 $\\
23-24 &$ 0.728\pm 0.034$&$ 0.590\pm 0.009$&$ 8.605\pm 3.850$&$0.018\pm0.004 $\\
24-25 &$ 0.711\pm 0.051$&$ 0.562\pm 0.017$&$ 7.866\pm 3.884$&$0.018\pm0.005 $\\
 \hline
   \multicolumn{5}{c}{$\log(\rho(M_\sun/pc^2))$}  \\ \hline
4.8-3.6 &$ 0.841\pm 0.010$&$ 0.765\pm 0.016$&$13.281\pm 1.242$&$0.029\pm0.003 $\\
3.6-3.2 &$ 0.822\pm 0.008$&$ 0.758\pm 0.010$&$11.262\pm 1.812$&$0.030\pm0.003 $\\
3.2-2.8 &$ 0.799\pm 0.005$&$ 0.729\pm 0.007$&$ 9.661\pm 1.187$&$0.028\pm0.002 $\\
2.8-2.4 &$ 0.776\pm 0.006$&$ 0.692\pm 0.006$&$ 9.052\pm 1.400$&$0.025\pm0.002 $\\
2.4-2.0 &$ 0.764\pm 0.009$&$ 0.656\pm 0.006$&$10.060\pm 1.746$&$0.021\pm0.002 $\\
2.0-1.6 &$ 0.733\pm 0.016$&$ 0.622\pm 0.005$&$ 7.499\pm 2.210$&$0.021\pm0.003 $\\
1.6-1.2 &$ 0.721\pm 0.034$&$ 0.584\pm 0.012$&$ 7.959\pm 3.654$&$0.018\pm0.004 $\\
1.2-0.8 &$ 0.721\pm 0.034$&$ 0.575\pm 0.017$&$ 8.118\pm 3.318$&$0.018\pm0.004 $\\
 \hline
   \multicolumn{5}{c}{Radial distance (kpc)} \\ \hline
0- 3 &$ 0.813\pm 0.018$&$ 0.746\pm 0.022$&$10.528\pm 2.882$&$0.030\pm0.005$\\
3- 7 &$ 0.773\pm 0.009$&$ 0.713\pm 0.007$&$ 6.571\pm 1.458$&$0.030\pm0.003$\\
7-15 &$ 0.793\pm 0.008$&$ 0.689\pm 0.006$&$13.500\pm 0.740$&$0.021\pm0.001$\\
15-25 &$ 0.776\pm 0.008$&$ 0.655\pm 0.007$&$12.488\pm 1.522$&$0.019\pm0.001$\\
25-35 &$ 0.760\pm 0.013$&$ 0.634\pm 0.006$&$10.784\pm 2.161$&$0.019\pm0.002$\\
35-55 &$ 0.734\pm 0.019$&$ 0.600\pm 0.005$&$ 8.674\pm 2.421$&$0.018\pm0.002$\\ 
55-85 &$ 0.691\pm 0.036$&$ 0.561\pm 0.011$&$ 5.864\pm 2.922$&$0.019\pm0.004$\\
85-120&$ 0.678\pm 0.052$&$ 0.569\pm 0.023$&$ 5.298\pm 3.536$&$0.023\pm0.008$\\

    \hline
   
 \end{tabular} 
  
 \caption{Colors, ages and metallicities with their corresponding errors of the cluster Abell 2744 as a function of: $\mu_J$, $\log(\rho)$ and radial distance. }\label{table1}

\end{table*}

\appendix

\section{Effect of the PSF}

In order to estimate a possible contamination of the colors in the outer regions of the cluster due to the
scattered light produced by the different band PSFs, we run the following tests. We have created an
artificial galaxy with an effective radius of $5$ kpc. This value is typical of galaxies with 10$^{11}$
M$_\sun$ \citep[see e.g.][]{Shen2003, Cebrian2014}. The mock galaxy follows a de Vaucouleurs $r^{1/4}$ profile \citep{deVaucouleurs1948}. The surface brightness profile of the simulated galaxy was convolved with the different HST PSFs. As we are interested on exploring the effects cause by the PSF wings in the outer region of our profiles, our PSFs
were retrieved from the HST PSF modeling software \textsc{tinytim}. We retrieved from the \textsc{tinytim}
webpage the largest PSF available for each band \footnote{Ideally, we would like to base our analysis using an empirical PSF. Unfortunately, there are not bright, isolated stars in the field of view to conduct such exercise.}. This allows us to use PSFs that extends at least over $12.5$\arcsec (i.e. $55$ kpc at the cluster redshift). We also simulated the effect of adding an ICL profile, an exponential profile with scale length of $18$ kpc and $\mu_J=24.5$ mag/arcsec$^2$ at R$=50$ kpc.

\begin{figure*}[!h]
 \begin{center}
  \includegraphics[scale=0.5]{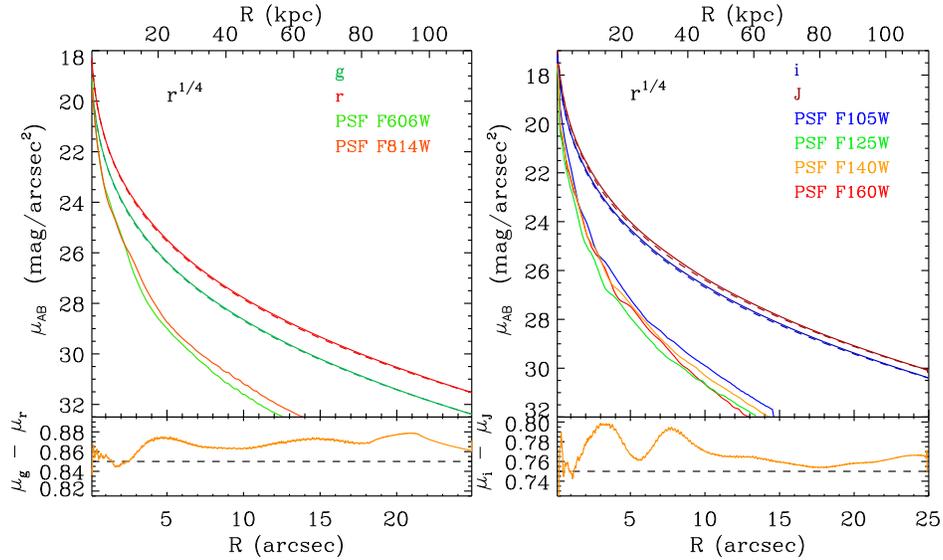}
  
 \caption{Upper panels: Surface brightness profiles of the models (dashed lines) and the convolved models
(solid lines) for the ACS (left hand panel) and WFC3 (right hand panel) filters used in the analysis. The
PSF profiles in all the observed used bands are also shown. The bottom panels show the effect of the PSF
in the restframe colors: $g-r$ (left) and $i-J$ (right). The input model colors are indicated with dashed
lines.} 

 \label{fig3}
  \end{center}
\end{figure*}

\begin{figure*}[!h]
 \begin{center}
  \includegraphics[scale=0.5]{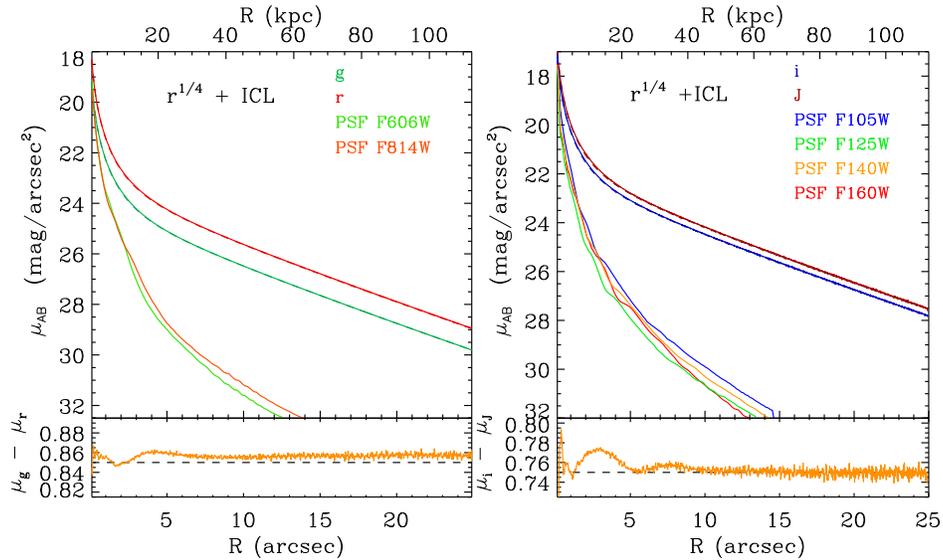}
  
 \caption{Same as in Fig \ref{fig3} but adding an ICL component of scale length $18$ kpc and $\mu_J=24.5$ mag/arcsec$^2$ at R$=50$ kpc.} 

 \label{fig4}
 \end{center}
\end{figure*}

The original model profiles and the convolved PSF models are shown in Fig. \ref{fig3} (without ICL) and \ref{fig4} (with ICL). As input colors for the model profiles we have used $g-r=0.85$ and $i-J= 0.75$, which are typical colors of the inner regions of the galaxies. In Fig. \ref{fig3}, we find that the effect of the different PSFs and their wings in the constructed radial colors is less than $0.03$ and $0.02$ mag in the $g-r$ and $i-J$ colors, respectively, at the region of the ICL ($R>50$ kpc). However, in Fig. \ref{fig4} where the ICL is included, we find that the smoother profile of the ICL diminishes the effect of the PSF and the colors differ in less than $0.01$ mag from the simulations. We warn the reader that these are the results of a simple simulation. We only consider the effect of one galaxy and the ICL region could be affected by the scattered light of other galaxies of the cluster. Consequently, the only purpose of these simulations is to provide a hint about how relevant could be the PSF effect on the ICL region.

%%%%%%%%%%%%%%%%

%%%%%%%%%%%%%%%%%%

\end{document}